\documentclass[runningheads]{llncs}
\usepackage[T1]{fontenc}
\usepackage{yhmath}
\usepackage{graphicx}
\begin{document}
\title{Variational Quantum Neural Networks (VQNNS) in Image Classification }
\author{Meghashrita Das\inst{1,2},
Tirupati Bolisetti\inst{1}}
%

\institute{University of Windsor, Windsor, Canada \and Indian Institute of Technology Kharagpur, Kharagpur, India\\
\email{meghashrita99@gmail.com},
\email{tirupati@uwindsor.ca}}
\maketitle              % typeset the header of the contribution
\begin{abstract}
Quantum machine learning has established as an interdisciplinary field to overcome  limitations of classical machine learning and neural networks. This is a field of research which can prove that quantum computers are able to solve problems with complex correlations between inputs that can be hard for classical computers. This suggests that learning models made on quantum computers may be more powerful for applications, potentially faster computation and better generalization on less data. The objective of this paper is to investigate how training of quantum neural network (QNNs) can be done using quantum optimization algorithms for improving the performance and time complexity of QNNs. A classical neural network can be partially quantized to create a hybrid quantum-classical neural network which is used mainly in classification and image recognition. In this paper, a QNN structure is made where a variational parameterized circuit is incorporated as an input layer named as Variational Quantum Neural Network (VQNNs). We encode the cost function of QNNs onto relative phases of a superposition state in the Hilbert space of the network parameters. The parameters are tuned with an iterative quantum approximate optimisation (QAOA) mixer and problem hamiltonians. VQNNs is experimented with MNIST digit recognition (less complex) and crack image classification datasets (more complex) which converges the computation in lesser time than QNN with decent training accuracy.  

\keywords{Quantum Neural Network \and Parameterized circuit \and Image classification \and Hilbert space \and Quantum Approximate Optimization }
\end{abstract}
\section{Introduction}
 Quantum computation offers potential avenue to increase the power of machine learning, neural networks models which is discussed in this paper with an algorithm. The objective of this research is to find a way to improve performance of QNN by variational approaches\cite{alam2022resilient}. We have used QAOA to train QNNs. Rather than using one qubit parameterized single circuit in input layer of QNN, we have used a 2 qubit parameterized circuit which is having Cost and Mixer Hamiltonians to calculate the cost of the classical QNN. QAOA uses a parameterized quantum circuit $\big|\Psi(\theta) \big\rangle$ to generate trial wave functions which is $\big|\Psi(\gamma,\beta) \big\rangle$ to compute or estimate the expectation value of $\big\langle \Psi(\gamma,\beta)|H|\Psi(\gamma,\beta) \big\rangle$ with respect to the problem Hamiltonian $H$. Now this variational QNN is applied on image processing applications (Classification and Recognition) which shows that VQNNs can improve time complexity and training accuracy\cite{liao2021quantum}.    

\section{Hybrid Quantum Classical Neural Networks}
Quantum machine learning (QML) proposes new types of models that leverage quantum computers’ unique capabilities to, for example, work in exponentially higher-dimensional feature spaces to improve the accuracy of models. A neural network is an elaborate function that is built by composing smaller building blocks called neurons\cite{oh2020tutorial}. A neuron is nonlinear function that maps one or more inputs to a single real number. Graphically, we represent neurons as nodes in a graph and each edge in our graph is associated with a scalar-value called a weight. The idea of QNN is to multiply each neuron by a different scalar before being collected and processed into a single value. The objective when training a neural network consists primarily of choosing our weights which is the building block oh hybrid QNNs\cite{wootton2021teaching}.

\section{Variational Quantum Neural Networks (VQNNs)}
\subsection{Parameterized quantum circuit}
A variational quantum circuit is comprised of three components. First, a feature map F which maps classical data point x into a m qubit quantum state $\big|\Psi\big\rangle$ :
\begin{equation}
    \big|\Psi(x) \big\rangle = F(x) \big|0\big\rangle ^ \otimes m
\end{equation}
Second, an ansatz which will build the quantum state with entanglements and rotation gates. The rotational angle of ansatz is parameterized by a vector $\theta$\cite{sebastianelli2021circuit}.
\begin{equation}
    \big|\phi(x,\theta) \big\rangle = A(\theta)\big|\Psi(x) \big\rangle  
\end{equation}
Finally, an observable $O$ is measured, and the eigenvalue corresponding to the resultant quantum state is recorded\cite{killoran2019continuous}. A variational quantum circuit is run repeatedly with an input $x$ and parameter vector $\theta$. The circuit’s expectation value (f) can be, 
\begin{equation}
    f(x,\theta) = \big\langle \phi \big( x, \theta \big) |O| \phi \big( x, \theta \big) \big\rangle  
\end{equation}
When a variational quantum circuit is used for machine
learning, this approximated expectation value is typically treated as the output of the model.

\subsection{QAOA}
QAOA is a hybrid quantum classical variational algorithm. Unlike the gate-based quantum computation model, it is based on adiabatic theorem from quantum mechanics. In this model, to perform any computation we need two Hamiltonians called H\textsubscript{M} and H\textsubscript{C}. Amongst them, the ground state of H\textsubscript{M} should be a preparable state and H\textsubscript{C} encodes the solution to our problem. QAOA is the trial a parameterized state $\big|\Psi(\gamma,\beta) \big\rangle$ (for some values of parameters $\gamma$, $\beta$) that should be prepared on a quantum computer. This state will give an expectation value of $\big\langle \Psi(\gamma,\beta)|H|\Psi(\gamma,\beta) \big\rangle$ with respect to the problem Hamiltonian $H$\cite{streif2019comparison}.

\subsection{VQNN Algorithm}
In this paper, VQNN is proposed where the quantum training process can be described as the state evolution with the Hilbert space of the parameter register and the QNN register\cite{streif2020training}. The quantum training protocol acts on both the parameter register ($\gamma$, $\beta$) and QNN register to encode the cost function of QNN which is a variant of the original QAOA Mixers\cite{arthur2022hybrid}. These operations can be mathematically expressed as $ e^{-i\gamma C\theta}$ and $ e^{-i \beta H\textsubscript{M}}$, where $\theta$ are the parameter vectors of QNN, $C(\theta$ ) is the cost function of the QNN, and $\gamma$i and $\beta$i are tunable hyperparameters. By heuristically tuning the hyperparameters, the quantum training can be done in on the optimal parameters of the QNN after iterations of the QAOA ansatz operations\cite{beer2021training}.
\subsubsection{QAOA ansatz}
\begin{figure}[!htb]
     \centering
     \includegraphics[width=.5\linewidth]{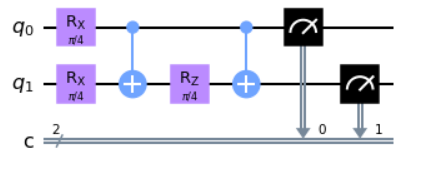}
     \caption{Quantum Parameterized Circuit}
     \label{fig:fig1}
\end{figure}

:
The QAOA ansatz has two parts, Mixer and Cost Hamiltonian. The Mixer Hamiltonian (H\textsubscript{M}) is choosen as Eq. \ref{eq:4},
\begin{equation}
\label{eq:4}
%\begin{split}
    H\textsubscript{M} = \sum_{j=1}^n X\textsubscript{j}
%\end{split}
\end{equation}
where X\textsubscript{j} is the Pauli X operator acting on the jth qubit. Then the Cost Hamiltonian (H\textsubscript{C}) is formed with rotational Pauli $z$ gate and cx gates with a rotational angle $\theta$. Finally a measurement in the computational basis is performed on the state\cite{mari2020transfer}. Repeating the above state preparation and measurement, the expected value of the cost function,
\begin{equation}
\label{eq:5}
%\begin{split}
   \big\langle C \big\rangle = \langle\big (\beta,\gamma)|H\textsubscript{C}|(\beta,\gamma) \big\rangle
%\end{split}
\end{equation}
Here the expectation of QAOA ansatz for rotaltional angle as $\pi$/4 will be -0.124. Fig.\ref{fig:fig1} represents the quantum circuit.

\subsubsection{Training with QNN}:
A Neural Network with 2 layers of ConvNN and a fully connected layer at the end is used and the value of the last neuron of the fully-connected layer is fed as the parameter $\theta$ into the above quantum QAOA ansatz circuit\cite{khairy2020learning}.

\section{Experimental Settings}
VQNN algorithm is experimented with MNIST digit datasets and crack detection datasets. MNIST image dataset is taken as a simple and relatively less complex where the performance of VQNN is measured and later it is experimented with a bit complex image recognition dataset to measure its compatibility. MNIST dataset contains 200 images from the training sample which contains hand written 0 and 1 digits. The crack detection kaggle dataset consists of various concrete surfaces labelled either negative if the surface presents no cracks and positive otherwise. Each class has 1,000 images for a total of 2,000 images.

\section{Results}
\begin{figure}[!htb]
   \begin{minipage}{0.47\textwidth}
     \centering
     \includegraphics[width=.9\linewidth]{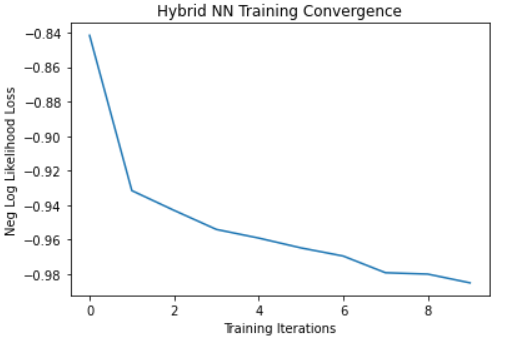}
     \caption{Training convergence of QNN on MNIST}\label{Fig:Fig2}
   \end{minipage}\hfill
   \begin{minipage}{0.47\textwidth}
     \centering
     \includegraphics[width=.9\linewidth]{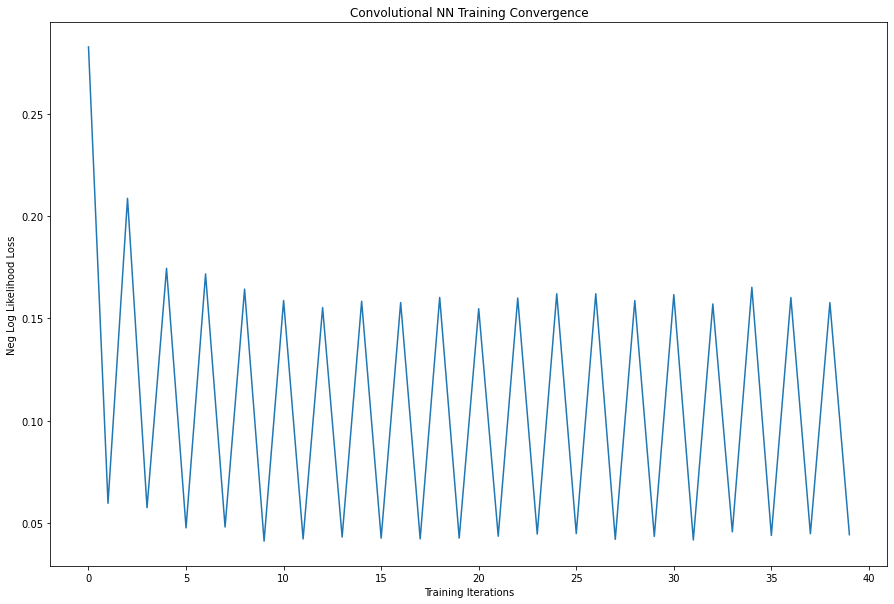}
     \caption{Training convergence of QNN on Crack detection}\label{Fig:Fig3}
   \end{minipage}
\end{figure}

\begin{figure}[!htb]
   \begin{minipage}{0.47\textwidth}
     \centering
     \includegraphics[width=.9\linewidth]{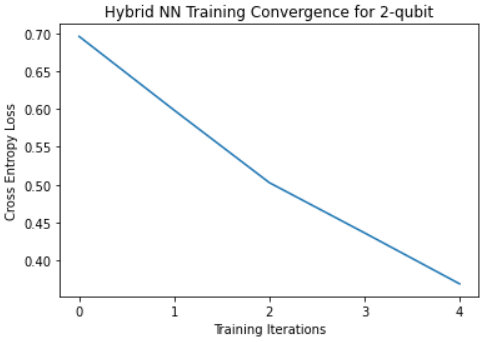}
     \caption{Training convergence of VQNN on MNIST}\label{Fig:Fig4}
   \end{minipage}\hfill
   \begin{minipage}{0.47\textwidth}
     \centering
     \includegraphics[width=.9\linewidth]{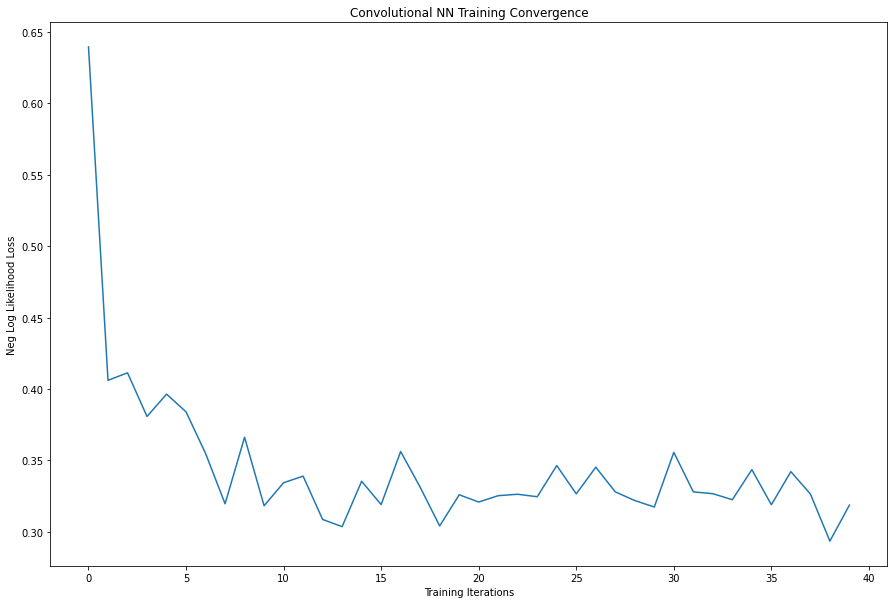}
     \caption{Training convergence of VQNN on Crack detection}\label{Fig:Fig5}
   \end{minipage}
\end{figure}

\begin{figure}[!htb]
   \begin{minipage}{0.47\textwidth}
     \centering
     \includegraphics[width=.6\linewidth]{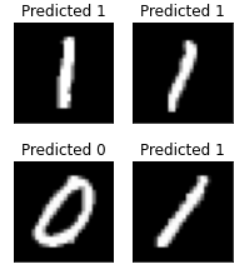}
     \caption{Prediction on MNIST for VQNNs}\label{Fig:Fig6}
   \end{minipage}\hfill
   \begin{minipage}{0.47\textwidth}
     \centering
     \includegraphics[width=.4\linewidth]{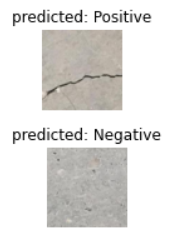}
     \caption{Prediction on Crack detection for VQNNs}\label{Fig:Fig7}
   \end{minipage}
\end{figure}

MNIST and Crack detection datasets are trained with QNN and VQNNs. The results are shown in the below Table~\ref{tab1} which shows that VQNN is taking less time than QNNs for the experiment with detecting surface cracks with an imporoved accuracy. Fig \ref{Fig:Fig2}, Fig \ref{Fig:Fig3}, Fig \ref{Fig:Fig4} and Fig \ref{Fig:Fig5} describe the training convergence for corresponding experiments with QNNs and VQNNs. Fig \ref{Fig:Fig6} is the predicted MNIST digits by VQNNs and Fig \ref{Fig:Fig7} is the predicted cracks as positive and negetive by VQNNs.  

\begin{table}
\caption{Results and Discussions}\label{tab1}
\begin{tabular}{|l|l|l|l|l|l|}
\hline
Index & Algorithm & Data-set & Training Accuracy & Validation Accuracy &Runtime\\
\hline
1 & QNN & MNIST & 99.90\% & 99.87\% & 1m 9 sec \\
2 & QNN & Crack detection & 87.05\% & 96.90\% & 48m 15 sec \\
3 & VQNN & MNIST & 99.88\% & 99.89\% & 8m 40 sec\\
4 & VQNN & Crack detection & 89.25\% & 89.00\% & 35m 48 sec \\
\hline
\end{tabular}
\end{table}

\section{Conclusion}
In this paper we proposed a framework to incorporate quantum optimisation routines for training QNNs. We have designed a QNN layes pipeline with QAOA ansatz tailored for training problems. In this research work, we have restricted ourselves up to a limited sized sample datasets. Also, we haven't incorporated any phase oracles in the ansatz circuit for better entanglement. So, our research contribution for this work can be to run VQNNs with a better efficient quantum back-end device where maximum qubits can be accomodated.

%
% the environments 'definition', 'lemma', 'proposition', 'corollary',
% 'remark', and 'example' are defined in the LLNCS documentclass as well.

\bibliographystyle{splncs04}
\bibliography{mybibliography}

\begin{thebibliography}{10}
\providecommand{\url}[1]{\texttt{#1}}
\providecommand{\urlprefix}{URL }
\providecommand{\doi}[1]{https://doi.org/#1}

\bibitem{alam2022resilient}
Alam, M.M.: Resilient quantum computing and machine learning  (2022)

\bibitem{arthur2022hybrid}
Arthur, D., et~al.: A hybrid quantum-classical neural network architecture for
  binary classification. arXiv preprint arXiv:2201.01820  (2022)

\bibitem{beer2021training}
Beer, K., List, D., M{\"u}ller, G., Osborne, T.J., Struckmann, C.: Training
  quantum neural networks on nisq devices. arXiv preprint arXiv:2104.06081
  (2021)

\bibitem{khairy2020learning}
Khairy, S., Shaydulin, R., Cincio, L., Alexeev, Y., Balaprakash, P.: Learning
  to optimize variational quantum circuits to solve combinatorial problems. In:
  Proceedings of the AAAI conference on artificial intelligence. vol.~34, pp.
  2367--2375 (2020)

\bibitem{killoran2019continuous}
Killoran, N., Bromley, T.R., Arrazola, J.M., Schuld, M., Quesada, N., Lloyd,
  S.: Continuous-variable quantum neural networks. Physical Review Research
  \textbf{1}(3),  033063 (2019)

\bibitem{liao2021quantum}
Liao, Y., Hsieh, M.H., Ferrie, C.: Quantum optimization for training quantum
  neural networks. arXiv preprint arXiv:2103.17047  (2021)

\bibitem{mari2020transfer}
Mari, A., Bromley, T.R., Izaac, J., Schuld, M., Killoran, N.: Transfer learning
  in hybrid classical-quantum neural networks. Quantum  \textbf{4}, ~340 (2020)

\bibitem{oh2020tutorial}
Oh, S., Choi, J., Kim, J.: A tutorial on quantum convolutional neural networks
  (qcnn). In: 2020 International Conference on Information and Communication
  Technology Convergence (ICTC). pp. 236--239. IEEE (2020)

\bibitem{sebastianelli2021circuit}
Sebastianelli, A., Zaidenberg, D.A., Spiller, D., Le~Saux, B., Ullo, S.L.: On
  circuit-based hybrid quantum neural networks for remote sensing imagery
  classification. IEEE Journal of Selected Topics in Applied Earth Observations
  and Remote Sensing  \textbf{15},  565--580 (2021)

\bibitem{streif2019comparison}
Streif, M., Leib, M.: Comparison of qaoa with quantum and simulated annealing.
  arXiv preprint arXiv:1901.01903  (2019)

\bibitem{streif2020training}
Streif, M., Leib, M.: Training the quantum approximate optimization algorithm
  without access to a quantum processing unit. Quantum Science and Technology
  \textbf{5}(3),  034008 (2020)

\bibitem{wootton2021teaching}
Wootton, J.R., Harkins, F., Bronn, N.T., Vazquez, A.C., Phan, A., Asfaw, A.T.:
  Teaching quantum computing with an interactive textbook. In: 2021 IEEE
  International Conference on Quantum Computing and Engineering (QCE). pp.
  385--391. IEEE (2021)

\end{thebibliography}
\end{document}